# ENRICO FERMI E ETTORE MAJORANA: CONTINUITÀ E RINNOVAMENTO NELL'INSEGNAMENTO DELLA FISICA TEORICA

*Alberto De Gregorio*
*Salvatore Esposito*

RIASSUNTO. Nel 1927 Enrico Fermi, appena nominato professore, tenne presso l'istituto di via Panisperna il suo primo corso di Fisica teorica. L'anno seguente ebbe tra i suoi allievi Edoardo Amaldi, Emilio Segrè e Ettore Majorana. Le lezioni del 1927-28, delle quali possiamo agevolmente ricostruire il contenuto, dovettero avere un forte impatto nella formazione di questi giovani allievi, che divennero presto attivi nelle ricerche condotte a Roma. In questa prospettiva, il caso di Majorana si rivela dei tre il più interessante: tra il 1933 e il 1936 egli progettò, in veste di libero docente, di tenere tre corsi avanzati di fisica, prima di essere nominato per «alta fama» professore di ruolo in Fisica teorica, nel 1937. Dall'analisi delle lezioni che egli tenne a Napoli nel 1938, possiamo concludere che Majorana in parte si rifece alle lezioni di Fermi che seguì da studente, in parte espose argomenti molto avanzati per l'epoca, che conferivano al corso un'impronta assai moderna. Molti di quegli argomenti di frontiera erano citati già nei programmi che Majorana aveva presentato come libero docente alcuni anni prima.

*Tristo è quel discepolo*
*che non avanza il suo maestro.*
Leonardo Da Vinci

**L'ambiente di via Panisperna**

Per iniziativa di Orso M. Corbino, direttore dell'Istituto fisico della Regia università di Roma e influente uomo politico, nel 1926 è messa a concorso dalla Facoltà di scienze dell'università di Roma la prima cattedra italiana di Fisica teorica, che andrà a Enrico Fermi appena venticinquenne. Enrico Persico, di un anno più grande di Fermi, è secondo e sarà chiamato poco dopo a Firenze. Aldo Pontremoli, terzo, andrà a Milano, ma nel 1928 perirà tragicamente, poco più che trentenne, nella spedizione del dirigibile «Italia» guidata da Umberto Nobile. Il giudizio della commissione è decisamente lusinghiero nei confronti di Fermi:

> La Commissione, esaminata la vasta e complessa opera scientifica del prof. Fermi, si è trovata unanime nel riconoscerne le qualità eccezionali e nel constatare che egli, pure in così giovane età e con pochi anni di lavoro scientifico, già onora altamente la fisica italiana. Mentre possiede in modo completo le più sottili risorse della matematica, sa farne uso sobrio e discreto, senza mai perdere di vista il problema fisico di cui cerca la soluzione e il giuoco e il valore concreto delle grandezze fisiche che egli maneggia. Mentre gli sono perfettamente familiari i concetti più delicati della meccanica e della fisica matematica classica, riesce a muoversi con piena padronanza nelle questioni più difficili della fisica teorica moderna, cosicché egli è oggi il più preparato e il più degno per rappresentare il nostro Paese in questo campo di così alta e febbrile attività scientifica mondiale. La Commissione pertanto [...] ritiene di poter fondare su lui le migliori speranze per l'affermazione e lo sviluppo futuro della fisica teorica in Italia.

L'istituto di via Panisperna aveva necessità di trasformarsi in una struttura moderna, del tipo di quelle esistenti all'estero. A Fermi, cui erano affidate le ricerche in campo teorico, fu affiancato per la parte sperimentale Franco Rasetti: entrambi, sotto la direzione di Corbino, costituiranno le colonne portanti della nascente scuola di fisica romana. La stima di Corbino traspare chiaramente:

> Nel Rasetti sono accoppiate con raro equilibrio le doti di un brillante e accurato sperimentatore con una profonda conoscenza della fisica teorica fin nei suoi più moderni sviluppi e con qualità di organizzatore di primo ordine. La sua competenza è universalmente riconosciuta in Italia e all'estero dove fu più volte invitato a tenere conferenze.

Quando Fermi, all'inizio del 1927, comincia a insegnare e a lavorare a Roma, soltanto una dozzina di studenti frequenta i quattro anni del corso di laurea in Fisica. La prima cosa da fare consiste, allora, nel reclutarne altri, possibilmente validi e motivati. Corbino se ne assume il compito, cominciando a tenere discorsi agli studenti del primo biennio di Ingegneria che frequentano il suo corso di Fisica sperimentale, esortando alcuni tra i migliori a mutare indirizzo di studi:

> Qualcuno di voi dovrebbe lasciare gli studi di ingegneria e passare a fisica, perché abbiamo qui a Roma un nuovo professore di fisica. Posso assicurarvi che questo è l'uomo capace di portare la fisica italiana ad un livello molto alto. In questo momento i giovani dovrebbero darsi alla fisica.

Edoardo Amaldi, Emilio Segrè e Ettore Majorana seguono l'invito di Corbino. L'ambiente che si va formando intorno a Fermi, e che porta all'affermarsi di una nuova scuola di fisica, è favorito senz'altro dalla piccola differenza di età tra maestri e discepoli ma anche, come ricorda Rasetti, dal fatto che le persone coinvolte sono poche:

> A parte il genio e la personalità di Fermi, una condizione indispensabile per il formarsi di un ambiente come la scuola di Roma ai tempi di Corbino era il numero esiguo di partecipanti.

Corbino promuove paternamente il lavoro dei giovani allievi, togliendo loro, come afferma Amaldi, «ogni preoccupazione per le finanze e l'amministrazione del laboratorio in modo che essi potessero serenamente dedicare tutto il loro tempo alla ricerca».

**I corsi di Fisica teorica**

In Italia, ancora negli anni Venti e per molti anni a venire, la tradizione risalente a Galileo Galilei imponeva che lo studio della fisica fosse sostanzialmente da intendersi come *sperimentale*, sebbene l'aspetto teorico non rimanesse del tutto escluso dalle discussioni scientifiche. Emblematico è il caso dello stesso Fermi che, come ricorda Segrè, conseguì la laurea in Fisica presso l'università di Pisa con una tesi che, «per tradizione, doveva essere di fisica sperimentale», mentre la tesi di diploma alla Scuola normale superiore fu di stampo fisico-matematico. Suscita dunque scarsa meraviglia il fatto che non vi fosse ancora in Italia un docente titolare di cattedra in Fisica teorica e che, ancora per diversi anni, quei fermenti della fisica quantistica che godevano di ampio spazio di discussione oltralpe non trovassero da noi accoglienza.

Il 20 gennaio 1927 Fermi tenne la prima lezione del corso di Fisica teorica. Gli appunti raccolti «dai Dott. Dei e Martinozzi» furono poi pubblicati in un libro, che si rivela perciò molto utile a comprendere l'approccio didattico adottato, sebbene uno sguardo ancora più approfondito nell'insegnamento della fisica teorica da parte di Fermi sia reso possibile dall'analisi di un altro suo libro, pubblicato nel settembre del 1928. I libretti relativi all'anno accademico 1927-28 dimostrano come gli argomenti illustrati a lezione siano gli stessi esposti nei primi sei capitoli del manuale. Per ciò che riguarda il secondo corso di Fisica teorica, dunque, pur non avendo a disposizione appunti presi durante le lezioni, possiamo ritenere di conoscere gli argomenti trattati grazie a tale libro e alla stretta corrispondenza tra il suo contenuto e ciò che Fermi registrò nel libretto delle lezioni.

Il corso del 1927-28, però, è particolarmente interessante anche perché fu quello al termine del quale Amaldi, Segrè e Majorana sostennero l'esame di Fisica teorica con Fermi, nel luglio del 1928. È evidente che gli argomenti trattati e lo stesso modo di esporli a lezione devono aver inciso profondamente sulla formazione scientifica di questi tre allievi. Amaldi e Majorana, per giunta, ebbero occasione di seguire un altro corso del loro professore di Fisica teorica, che l'anno seguente, 1928-29, tenne anche il corso di Fisica terrestre. Fu l'ultimo esame sostenuto da Amaldi e Majorana, che pochi giorni dopo, il 6 luglio 1929, si laurearono in Fisica.

**Il programma e il libro**

In questo paragrafo ci soffermeremo più in dettaglio sugli argomenti delle lezioni e del libro «Introduzione alla fisica atomica», fornendo alcune indicazioni utili a comprendere quanto fedelmente seguano uno sviluppo comune a entrambi.

Per primo ci soffermiamo sul libretto delle lezioni relativo al corso di Fisica teorica del 1927-28. Vi sono registrate sessantacinque lezioni, che vanno dal 15 novembre 1927 al 2 giugno 1928 e che è facile raccogliere in sei gruppi, in base agli argomenti trattati. Un primo gruppo riguarda la teoria cinetica dei gas e gli elementi della meccanica statistica. Fermi affronta qui argomenti classici, come la dipendenza della pressione dall'energia cinetica delle molecole, i cammini liberi medi, l'equipartizione dell'energia, le distribuzioni di Boltzmann e di Maxwell. Un secondo gruppo comprende dieci lezioni dedicate all'elettromagnetismo, in cui ad esempio Fermi illustra il significato delle perturbazioni elettromagnetiche, del vettore di Poynting, della teoria elettronica delle dispersione e della teoria della radiazione. Abbiamo poi un terzo gruppo, sui corpuscoli costituenti la materia: gli elettroni, le trasformazioni radioattive, i raggi alfa, beta e gamma. Nel quarto gruppo Fermi espone i fondamenti della teoria atomica; i principali argomenti sono il modello atomico di Rutherford, i quanti di luce, l'effetto Compton. Il 'nucleo' del corso tenuto da Fermi è rappresentato però dall'atomo di Bohr; ben venti delle sessantacinque lezioni trattano questo argomento: dai livelli energetici alla costante di Rydberg, dalle condizioni di Sommerfeld al principio di corrispondenza, dalla struttura fine allo *spin*. Il sesto gruppo di lezioni riguarda infine gli spettri degli atomi con uno, due o tre elettroni di valenza.

Se passiamo adesso a considerare l'indice della «Introduzione alla fisica atomica», la corrispondenza tra argomenti è evidente. Nell'ordine, i titoli dei primi sei capitoli del libro sono: *La teoria cinetica dei gas*, *Teoria elettromagnetica della luce*, *I corpuscoli elettrici*, *Gli scambi energetici tra luce e materia*, *L'atomo di Bohr*, *Le molteplicità spettrali*. Possiamo senz'altro ritenere che la stesura del manuale di fisica atomica e l'esposizione degli argomenti durante il corso si siano influenzate reciprocamente in maniera molto stretta, e forse spingerci fino ad avanzare persino l'ipotesi che Fermi abbia organizzato l'indice e il programma in modo contestuale.

**Ettore Majorana: da allievo a professore**
La traccia lasciata dai corsi di Fermi su uno degli allievi "teorici", Ettore Majorana, è riscontrabile nei quaderni di appunti personali di quest'ultimo, soprattutto nei cosiddetti *Volumetti* (cinque quaderni di appunti personali redatti tra il 1927 e il 1932 circa, sui quali Majorana riportò alcuni argomenti di studio). Se non vi compare alcun cenno a questioni e fatti di natura specificatamente sperimentale, è comunque illuminante dare anche solamente uno sguardo all'indice dei temi trattati in essi, per ritrovarvi diversi problemi che, introdotti proprio nei corsi di Fermi – sia in quello di Fisica teorica sia in quello di Fisica terrestre –, Majorana rielaborò autonomamente da studente. Come esempio citeremo alcuni titoli soltanto: *Quantizzazione dell'oscillatore lineare armonico*, *Statistica e termodinamica*, e anche *Equilibrio di una massa liquida eterogenea in rotazione* (*Problema di Clairaut*). Tale abitudine di Majorana di annotare nei volumetti i problemi affrontati e risolti, o di riportare quelli in fase di risoluzione nei cosiddetti *Quaderni* (un'altra ventina di fascicoli, il cui contenuto è organizzato in modo meno strutturato rispetto ai *Volumetti*) non rimarrà confinata agli anni in cui egli frequentava l'università da studente, ma si estenderà ben oltre e indurrà il fisico catanese a registrare in maniera più o meno regolare le ricerche svolte, delle quali, altrimenti, in molti casi non sarebbe rimasta alcuna traccia, neppure nei ricordi sbiaditi di colleghi e amici.

Nel 1933 Majorana trascorse, per motivi di studio e ricerca, alcuni mesi a Lipsia, dove entrò in contatto con Werner Heisenberg. In occasione di un breve periodo di vacanza, in primavera tornò a Roma. Le biografie riferiscono che dopo aver concluso il soggiorno in Germania Majorana si appartò sempre più dalla vita dell'Istituto, imboccando quel percorso di progressivo isolamento personale e intellettuale che lo avrebbe portato, nel 1938, a scomparire misteriosamente. Eppure, nel maggio del 1933 egli propose di tenere, in qualità di libero docente, un corso di *Metodi matematici della meccanica quantistica*, del quale resta però solamente il programma che egli presentò preventivamente. Allo stesso modo, nella primavera del 1935 egli consegnò il programma per un corso di *Metodi matematici della fisica atomica* e l'anno successivo per uno di *Elettrodinamica quantistica.*

Nel primo di questi tre programmi, a una trattazione fondata sulle vecchie regole di Bohr e Sommerfeld o sulla meccanica di Schrödinger e di Heisenberg, Majorana preferì la teoria dei gruppi, con cui si proponeva di illustrare sia il mondo dei quanti sia i risultati fondamentali della relatività. Più marginale il ruolo riservato alla teoria dei gruppi nel programma del 1935-36, in cui l'aspetto fenomenologico della fisica atomica assumeva un peso maggiore. Nell'ultimo programma, infine, trovano spazio rielaborazioni e contributi originali all'elettrodinamica quantistica, tra i quali spicca senz'altro la già ricordata «simmetria delle cariche».

Majorana non tenne mai i corsi per i quali presentò questi tre programmi, forse per mancanza di studenti, ma queste seppur minime iniziative suggeriscono che non rinunciò mai completamente alla fisica, nemmeno in quelli che sono stati descritti come i suoi "anni bui".

La nomina in ruolo avvenne alla fine del 1937, «per alta fama di singolare perizia»:

> La commissione esita ad applicare a lui la procedura normale dei concorsi universitari. […] Fin dall'inizio della sua carriera scientifica ha dimostrato una profondità di pensiero e una genialità di concezioni da attirare su di lui la attenzione egli studiosi di Fisica Teorica di tutto il mondo. [I suoi lavori sono] tutti notevolissimi per l'originalità dei metodi impiegati e per l'importanza dei risultati raggiunti.

Nel gennaio del 1938 Majorana diede inizio presso l'Università di Napoli al suo primo corso universitario, in Fisica teorica, disciplina in cui Fermi gli era stato maestro.

**L'impronta del maestro**

La vecchia teoria dei quanti, nata principalmente dai contributi di Bohr e Sommerfeld nella prima metà degli anni Dieci del Novecento, imponeva alcune restrizioni – o per meglio dire discontinuità – al movimento degli elettroni intorno al nucleo e rappresentava, in definitiva, un compromesso che garantiva la massima adesione ai risultati sperimentali con il minimo scostamento dai principi della meccanica classica. La meccanica quantistica formulata a metà degli anni Venti da Schrödinger e Heisenberg era invece una teoria organica mediante la quale rappresentare il mondo degli atomi, ma a tale aspetto di completezza finiva per sacrificare le rappresentazioni intuitive della realtà microscopica. Quest'ultima sua caratteristica non ne faceva materia di agile insegnamento nelle università, tant'è che, ancora per circa tre decenni, la «vecchia teoria dei quanti» conserverà un ruolo non secondario nell'insegnamento della fisica atomica.

Fatta una simile premessa, non stupisce che Fermi abbia escluso dal corso da lui tenuto nel 1927-28 l'insegnamento della nuova meccanica quantistica, limitandosi a riservargli il decimo capitolo del suo libro. Majorana stesso, dieci anni dopo, dedicò alcune lezioni alla vecchia teoria di Bohr e Sommerfeld, come possiamo dedurre da alcuni appunti autografi e dalla trascrizione fattane da un suo studente (tali appunti però non comprendono le prime quattro lezioni).

Un aspetto davvero interessante che emerge dalle circa dieci lezioni che Majorana dedicò alla vecchia teoria quantistica dell'atomo è la loro strettissima analogia con gli argomenti trattati a lezione nel 1927-28 da Fermi. Si tratta innanzitutto di un'analogia nei contenuti, ripresi, pur se in una trattazione meno articolata (circa dieci lezioni), quasi integralmente dal corso di Fermi. Ma è anche un'analogia che si spinge spesso fino a riguardare gli aspetti più minuti e formali (le formule e gli esempi portati, i simboli scelti ecc.).

L'effetto Compton, la struttura fine degli spettri, lo *spin* dell'elettrone, gli spettri dei metalli alcalini e alcalino-terrosi sono alcuni degli argomenti che Majorana mutuò in modo inequivocabile dalla «Introduzione alla fisica atomica» di Fermi, dedicando, in ultima analisi, ad argomenti che aveva seguito a lezione nel 1927-28 parte del proprio corso di Fisica teorica.

**Contributi originali**

Una volta discusse le applicazioni della vecchia teoria dei quanti, Majorana si dedicò all'elettromagnetismo e alla relatività e, conclusesi le vacanze per il carnevale, intraprese lo studio della fisica microscopica in termini della nuova meccanica quantistica, in entrambe le sue formulazioni: quella matriciale di Heisenberg e quella ondulatoria di Schrödinger. In questo modo conferì al proprio corso, poi purtroppo interrotto, connotazione di profonda modernità, che ha conservato fino ad oggi, a distanza di quasi settanta anni. Come abbiamo già accennato, nei corsi universitari la meccanica quantistica non aveva ancora assunto quel ruolo di completa autonomia dalla vecchia teoria dei quanti che avrebbe raggiunto in seguito; tanto più nella versione matriciale di Heisenberg, al cui insegnamento Majorana si dedicò invece con passione. Nel suo corso, dunque, può essere isolato un gruppo di lezioni dedicate a temi molto più avanzati che non quelli della vecchia teoria dei quanti. Se poi torniamo a considerare i tre programmi che presentò come libero docente tra il 1933 e il 1936, notiamo che già da allora Majorana aveva progettato di tenere lezione su tali argomenti, e in maniera ancora più articolata di quanto non fece poi di fatto nel 1938; ma, ancor più interessante, vi si trovano citati argomenti all'avanguardia persino come temi di ricerca, prima ancora che di insegnamento. Per portare un esempio, nel programma del 1936 Majorana incluse la «simmetria delle cariche», ossia la ricerca di una equazione quanto-meccanica relativistica, del tipo di Dirac, che fosse simmetrica rispetto all'elettrone e alla sua antiparticella, il positrone, da poco osservata sperimentalmente. Su tale questione egli avrebbe pubblicato, solamente nel 1937 e su sollecitazione di Fermi, un articolo in cui introduceva un argomento ancora oggi al vaglio della verifica sperimentale: la teoria dei *neutrini di Majorana*.

Un altro esempio, forse ancora più illuminante, ci è offerto dall'uso che Majorana prevedeva di fare della teoria dei gruppi, come traspare chiaramente dai programmi che presentò a Roma. Tale teoria si differenziava radicalmente dall'approccio matematico familiare ai fisici, e la maggior parte di essi l'avrebbe trascurata per almeno altri vent'anni. Introdotta nel quadro della meccanica quantistica nel 1928 attraverso il fondamentale libro di Hermann Weyl, fu, diversamente che per Fermi, uno degli interessi principali di Majorana, come è evidente dai suoi manoscritti. Questo 'incontro' con i nuovi e potenti mezzi matematici segnerà pressoché tutta la sua produzione successiva anche inedita, compreso l'approccio d'avanguardia che caratterizzerà il suo corso di Fisica teorica di Napoli.

**Conclusioni**

Quanto qui esposto dimostra innanzitutto quanto sia importante e utile l'analisi dei documenti originali nel condurre studi di carattere storico. Grazie a carte conservate in polverosi scaffali è stato possibile sia ricostruire il contenuto del corso di Fisica teorica di Enrico Fermi seguito da Ettore Majorana, sia constatare come l'interesse di quest'ultimo per l'insegnamento universitario risalisse a prima dell'unico corso che effettivamente tenne, a Napoli, nel 1938.

L'aspetto più interessante che emerge da tale analisi è senz'altro che l'attività didattica di Majorana, sia effettiva (il corso di Napoli) sia semplicemente in programma (i corsi da libero docente mai realmente tenuti), fu una sintesi tra l'esperienza maturata da studente e le successive rielaborazioni personali, in un delicato equilibrio tra continuità con la tradizione accademica della propria formazione e soluzioni originali apportatrici di rinnovamento.

# BIBLIOGRAFIA

Alberto De Gregorio, fisico, svolge attività di ricerca in storia della fisica presso l'Università di Roma «La Sapienza.

Salvatore Esposito è un fisico teorico dell'Università di Napoli «Federico II» che da alcuni anni lavora anche su questioni di storia della fisica.